# Embedding a carbon nanotube across the diameter of a solid state nanopore

E.S. Sadki[a)], S. Garaj, D. Vlassarev, J.A. Golovchenko, and D. Branton [b)]

Harvard University, Department of Physics, 17 Oxford Street, Cambridge, MA 02138 USA

[a)]Present address: Physics Department, Faculty of Science, United Arab Emirates University, Al Ain, P.O. Box 17551, Abu Dhabi, UAE

[b)]Harvard University, Department of Molecular and Cellular Biology, Jefferson 254, 17 Oxford Street, Cambridge, MA 02138 USA. Electronic mail: dbranton@harvard.edu

A fabrication method for positioning and embedding a single-walled carbon nanotube (SWNT) across the diameter of a solid state nanopore is presented. Chemical vapor deposition (CVD) is used to grow SWNTs over arrays of focused ion beam (FIB) milled pores in a thin silicon nitride membrane. This typically yields at least one pore whose diameter is centrally crossed by a SWNT. The final diameter of the FIB pore is adjusted to create a nanopore of any desired diameter by atomic layer deposition (ALD), simultaneously embedding and insulating the SWNT everywhere but in the region that crosses the diameter of the final nanopore, where it remains pristine and bare. This nanotube-articulated nanopore is an important step towards the realization of a new type of detector for biomolecule sensing and electronic characterization, including DNA sequencing.

# I.  INTRODUCTION AND RATIONALE

A nanopore through a solid state membrane or a biological cell membrane separating two ionic-solution filled compartments has shown itself to be a promising and versatile type of single molecule detector, capable of sensing the properties of both single small molecules and long polymers in solution.[1-7]  When a bias of 60 – 500 mV is applied across the membrane separating the two ionic solutions, charged polymeric molecules (e.g. DNA) in the solution are electrophoretically driven through the nanopore and each molecule can be detected as it traverses through the nanopore constriction.

Until now, molecular detection has usually relied on observing a drop in the ionic conductivity as each molecule is driven through a single nanopore.  The resolution and sensitivity of this molecule detector have been limited by the very small changes in ionic current and the high speed at which a DNA polymer translocates through the nanopore (usually >1 nucleotide/μsec for solid state nanopores).[8,9]  One is therefore led to consider other molecule detection modes that are capable of discerning structure along the molecule at higher resolution and greater electronic sensitivity.  Two possibilities are to sense molecules with a nanoscale field effect transistor (FET) embedded across the pore diameter or by observing tunneling or other forms of electronic transport between gapped electrodes embedded in the nanopore.[6,10]  Important steps to fabricating gapped metallic electrodes atop a nanopore have recently been demonstrated.[11,12]

But instead of metallic probes, an electrically contacted carbon nanotube across the diameter of the nanopore is particularly attractive because of its chemical stability, nanoscale dimensions and highly ordered structure.  Such a nanotube can serve as a FET[13,14] or, when suitably gapped, as the probes of a tunneling device.[6,15-17] During molecular transport through such a device, electronic rather than ionic currents can be sensed, potentially offering improved signal to noise and a more controlled environment for molecular transport. But producing a nanopore with an embedded single walled carbon nanotube across its diameter is a significant challenge that has not previously been confronted. Controlled positioning at specific device locations remains one of the most significant hurdles in the application of carbon nanotubes[18,19], and our goal was to devise a straightforward method that could easily, but reproducibly, produce many devices having an embedded single walled carbon nanotube across at least one nanopore of any chosen diameter. Here we demonstrate one way such devices can be realized.

## II. EXPERIMENTAL

Figure 1 presents a schematic of the fabrication method. First, a row of 1μm x 1μm iron (Fe) catalyst pads for CVD growth of carbon nanotubes were patterned on a low stress silicon nitride ($SiN_x$) membrane by electron beam lithography (EBL), electron beam evaporation, and metal lift-off. Then, rows of ~30nm diameter pores were milled through the membrane, usually in 11 x 11 arrays in the proximity of each of the 1μm x 1μm Fe pads with an FIB (Fig. 2a). The pores in each array were spaced 100 nm x 100 nm center-to-center, such that each array of 121 pores measured ~1μm x ~1μm edge-to-edge. Carbon nanotubes were then grown from the Fe pads by CVD.[20] Typically at least one of the carbon nanotubes growing from the Fe pad reached the pore array and crossed the center of at least one of the FIB pores. Because the Fe catalyst pad seeded the growth of multiple nanotubes that grew in many directions, we often noted several nanotubes that grew across other pores in the pore array.

To adjust the nanopore size and embed the nanotube in the finished nanopore, the samples were transferred to an ALD chamber that was used to conformally deposit multiple layers of insulating $Al_2O_3$ on all exposed surfaces, including the perimeter of the nanopores, but not on the portion of the nanotube that is to serve as an FET sensor or tunneling probe. ALD of high- dielectric materials has been shown to produce a benign dielectric/SWNT interface that does not adversely affect the electrical properties of the nanotube.[21,22] Because ALD of $Al_2O_3$ is a self-limiting process that depends on half-cycle reactions that proceed through the formation of Al–O Lewis acid-base complexes,[23] the hydrophobic surface of a pristine defect-free carbon nanotube, in which all the carbon atoms are in $sp^2$ configuration, does not initiate the growth of aluminum oxide from its surface.[22,24,25] Rather, the $Al_2O_3$ layers grew from all the silicon nitride surfaces, with lateral growth from the newly formed $Al_2O_3$ eventually burying only those portions of the carbon nanotube lying on the silicon nitride surface and near the inside diameter of the FIB pore. Deposition of a known number of $Al_2O_3$ layers, each of which is ~0.1 nm thick, on the inside diameter of the FIB pore shrank the pore diameter to a predictable smaller size,[26] with the length of exposed nanotube shrinking commensurately. Only the portion of the suspended carbon nanotube that passed across the finished ALD coated nanopore remained uncoated (Figure 3). Thus the $Al_2O_3$ coating could serve as a protective layer that insulates the nanotube from conducting fluid and the influence of proximal molecules anywhere in solution except in the finished nanopore.

We found the following procedural details consistently produced 2–10 nm nanopores whose diameters were crossed by one single-walled carbon nanotube. We started with 80 x 80 μm free-standing, low-stress 250-270 nm thick $SiN_x$ membranes, obtained by anisotropic KOH etching of a $SiN_x$ coated silicon wafer.[27] To assure cleanliness, the silicon chips bearing the membranes were cleaned by dipping in trichloroethylene, acetone, and finally methanol, and baked at 150°C in air. Subsequently, a row of 0.5 to 1 nm thick 1μm x 1μm Fe pads were patterned on the membrane using standard EBL processing (Raith-150 system) with a Poly(methyl methacrylate) (PMMA) resist that was lifted off with acetone followed by a rinse in isopropyl alcohol and drying in a $N_2$ atmosphere. The pore arrays were milled using a FEI Micrion 9500 FIB system operated at 50 kV acceleration voltage, with an aperture diameter of 15 μm corresponding to a 1.4 pA beam current. To obtain the smallest possible diameters of the pores, a single point (1 pixel) ion beam was rastered through the array ~100 times to obtain a total beam time of 6s/pore. The resulting pores typically had ~35 – 40 nm diameters and were spaced 100 nm apart in a square array. Each array, with typically 11 x 11 pores, was located 1 μm away from the pre-made Fe pads. Figure 2(a), shows a transmission electron microscope (TEM) image of an array of FIB milled pores (JEOL 2100 operated at 200 KV). In this example, the pores have a diameter of ~40 nm and it can be seen that the centered 11 pore x 11 pore array is but one of several similar pore arrays which were patterned in a row parallel to the row of pre-made Fe catalyst pads (not shown). It is noted that the area of the membrane between the pores is partially milled because the ion beam is not blanked as it moves from one point to the next.

Carbon nanotubes were grown from the Fe pads in a 1 inch diameter quartz tube furnace (Lindberg/Blue Mini-Mite 1100 C), with methane gas ($CH_4$) as the carbon source. First, the sample temperature was ramped to $900^oC$ in a flow of 500 sccm of Ar. When it reached $900^oC$, the sample was annealed in 200 sccm of pure $H_2$ for 10 mins. Carbon nanotube growth was initiated by introducing a 1000 sccm flow of $CH_4$ for 15 minutes together with the 200 sccm $H_2$ flow. After growth, the system was cooled down to room temperature in a 500 sccm flow of Ar. These conditions usually yielded the growth of single-walled carbon nanotubes that were several microns in length with a high probability of at least one nanotube from each Fe catalyst pad growing across a nanopore array. Through many trials, we found the distances between the Fe-pads and the pore arrays and the spacing between the pores in each array stated above yielded a high probability that a nanotube would centrally cross the diameter of at least one pore (Figure 2b).

Atomic layer deposition of $Al_2O_3$ was conducted in a home-built system at $225^oC$, by alternating cycles of flowing trimethylaluminum (TMA) vapor followed by water vapor. At the end of each cycle, one atomic layer of $Al_2O_3$ (thickness ~0.1 nm) was deposited.[5] As shown in Figure 3b (arrows), the interface between the initial pore perimeter and the deposited $Al_2O_3$ layer can be seen in electron micrographs and it is clear that the carbon nanotube over the finished nanopore remained uncoated after the ALD had shrunk the pore to the finished desired diameter. In this example, the final nanopore diameter was 11 nm, the added alumina thickness was 10 nm, and the nanotube diameter was 2.6 nm, but by using fewer or more ALD cycles, the same procedure was used to produce many other larger or smaller diameter nanopores crossed by an embedded nanotube. To keep the nanotube across the finished nanopore free from alumina deposition, direct exposure to an electron beam during SEM or TEM observation (as was done to produce the image in Figure 2b) must be avoided before the ALD processing. This is because direct exposure to an e-beam inevitably contaminates or causes defects in the nanotube which then nucleate alumina deposition.

With appropriate electrical connections to the carbon nanotube, the exposed length of a semiconducting nanotube that crosses the finished nanopore diameter (Figure 3b) can serve as an FET single molecule detector because the translocation of a molecule in its immediate vicinity will influence the carrier concentrations in the nanotube. Alternatively, a gapped nanotube with ends abutting the perimeter of the finished nanopore can serve as a tunneling device to sense and characterize molecules as they pass through the nanopore. To achieve the latter configuration, the uncovered portion of the nanotube crossing the nanopore (Figure 3b) can easily be eliminated by exposure to a low energy (e.g. 0.5-1 keV) sputter ion beam or plasma etch apparatus, which will rapidly remove the exposed length of the nanotube while leaving the alumina-covered portion of the original nanotube untouched. For illustrative purposes, even the electron beam of the TEM itself can be used, as can be seen in Figure 3c. But a low energy ion beam is preferred because its very short penetration distance into the alumina will minimize damage to the alumina covered portion of the gapped nanotube. Other methods of removing the suspended part of the nanotube over the nanopore, such as reactive ion etching, are currently being investigated.

## III. DISCUSSION AND CONCLUSION

Using an ungapped or gapped carbon nanotube integrated with a nanopore will require electrically connecting the appropriate positions of a selected nanotube to the outside world before they are buried and insulated by ALD. This can be accomplished without device damage or contamination by ice lithography[28] before the ALD procedure. As noted above, we have found that carbon nanotube exposure to an electron beam during standard EBL and lift-off damages or contaminates the nanotube. An unwanted consequence of such contamination or damage is that subsequent ALD coats the entire length of the nanotube,[25] including the portion that must remain uncoated if it is to serve as an FET or be gapped by an ion beam in the finished nanopore. Ice lithography avoids this problem because imaging, locating, and selecting the desired nanotube and determining the desired pattern of metal contacts are accomplished while the entire device, including the nanotube, is protected under a thin layer of water ice.[29] Furthermore, we and others have found that ALD of $Al_2O_3$ and other high-dielectric materials produces a benign dielectric/SWNT interface that does not adversely affect the electrical properties of the nanotube.[22,30] Finally, we note that if the electrical signals from only the contacted nanotube that successfully crosses the diameter of a single nanopore is used to sense the presence of molecules translocating through that nanopore, the presence of unused nanopores in the pore array is of no concern.

In summary, we have demonstrated a simple and reproducible nanoscale fabrication technique for articulating a nanopore with a carbon nanotube. Although producing arrays containing a large number of such articulate nanopores cannot depend on individually selected e-beam lithographic processing and will require more elaborate, controlled methods to position nanotubes across the diameter of multiple nanopores, the simple approach we have demonstrated here is an important step towards the realization of a nanodevice platform whose value for in-solution sensing and characterization of molecules and biopolymers needs to be established before more elaborate, highly parallel and better controlled fabrication methods are implemented.

## ACKNOWLEDGMENTS

This work was supported by NIH award number R01HG003703 to J. Golovchenko and D. Branton.

# Figure Captions

Figure 1. (Color online) Schematic diagram of the fabrication process of a carbon nanotube articulated nanopore. **(a)** Deposition of iron (Fe) pads on a silicon nitride membrane by a combination of electron beam lithography patterning and metal lift-off. **(b)** Arrays of nanopores are milled through the $SiN_x$ membrane by a FIB machine. **(c)** Chemical vapor deposition growth of a carbon nanotube from the Fe pad over the array of pores. **(d)** Zoom-in view of a nanopore from the array with a carbon nanotube crossing its diameter. **(e)** A layer of aluminum oxide ($Al_2O_3$) is deposited by ALD. **(f)** Zoom-in view of the carbon nanotube articulated nanopore after ALD showing the decrease of its diameter by ALD growth of $Al_2O_3$ from the pores initial perimeter.

Figure 2. Transmission electron microscopy images of **(a)** an FIB milled array of pores drilled through a $SiN_x$ membrane. **(b)** A FIB milled pore from an array after a single-walled carbon nanotube has grown across the pore's diameter. Scale bar is 10 nm.

Figure 3. Transmission electron microscopy images. **(a)** A nanopore array after the deposition of $Al_2O_3$ by ALD. The circular pattern of dots here and in b and c denotes the boundary of the initial FIB pore perimeter which encloses the subsequently deposited $Al_2O_3$ layers. **(b)** TEM image of a nanotube articulated nanopore after ALD deposition. **(c)** TEM image of the same nanopore as in b, after the portion of the nanotube not covered with $Al_2O_3$ has been ablated with the microscope's electron beam.

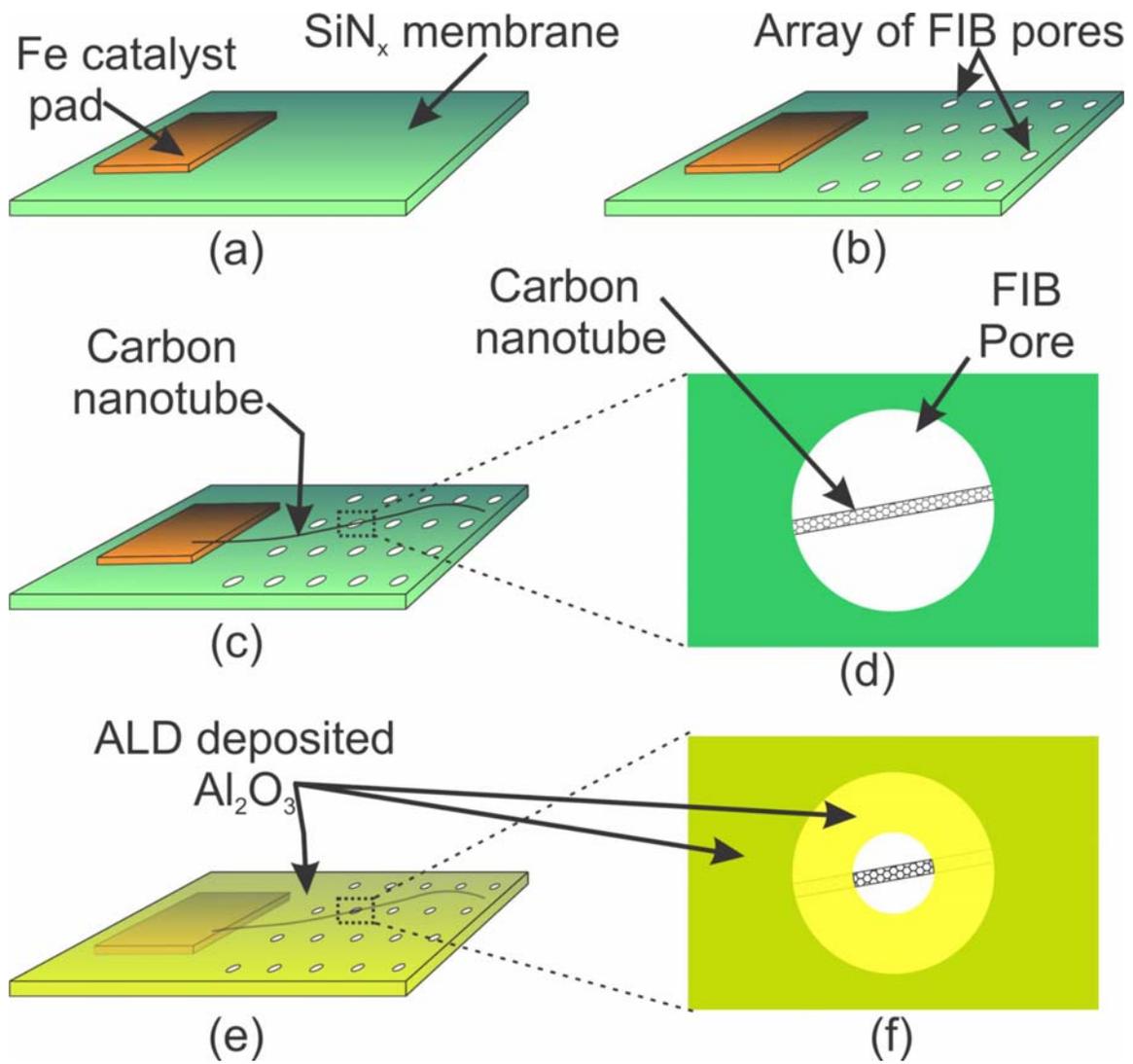

**Figure 1**

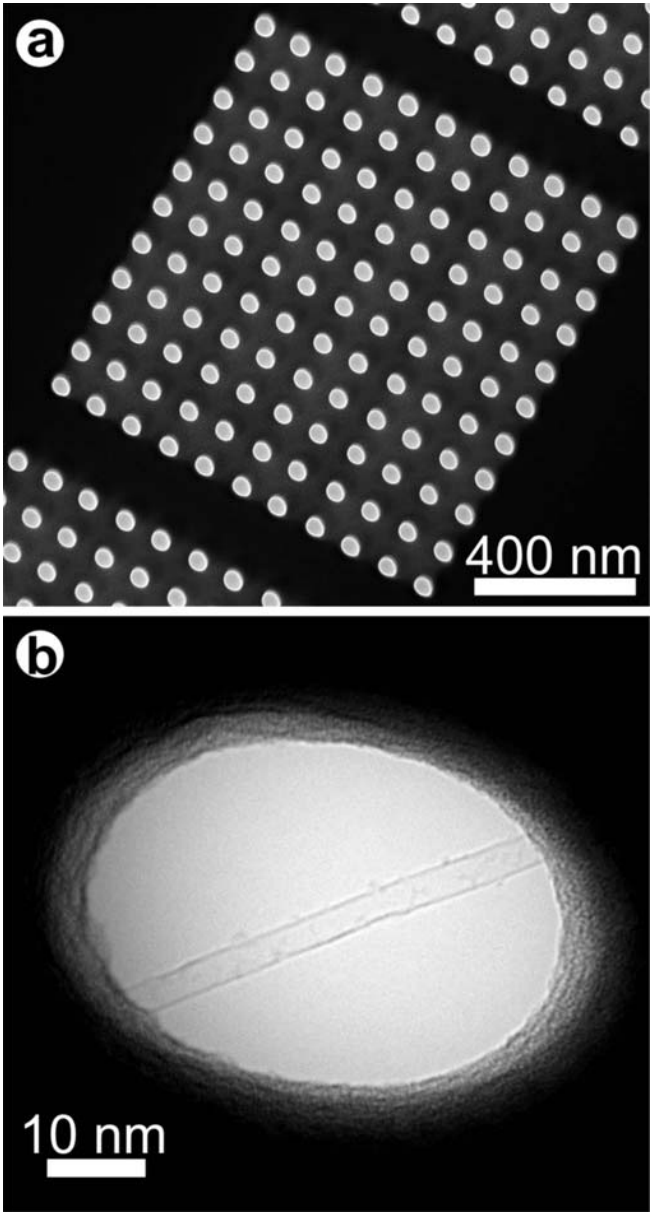

**Figure 2**

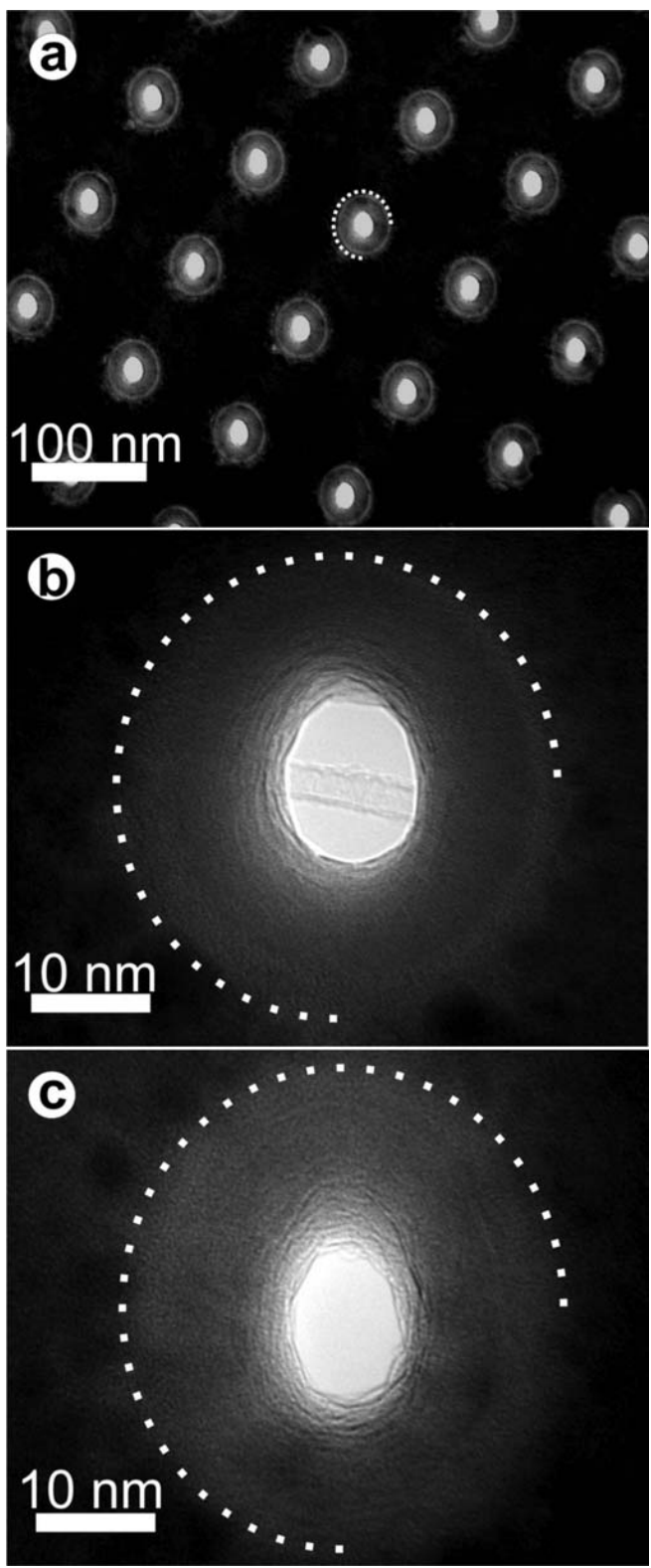

**Figure 3**